\documentclass{article}

\usepackage{arxiv}

\usepackage[utf8]{inputenc} 
\usepackage[T1]{fontenc}    
\usepackage{hyperref}       
\usepackage{url}            
\usepackage{booktabs}       
\usepackage{amsfonts}       
\usepackage{nicefrac}       
\usepackage{microtype}      
\usepackage{lipsum}
\usepackage{amsmath}

\usepackage{graphicx}
\title{Addressing Limited Weight Resolution in a Fully Optical
Neuromorphic Reservoir Computing Readout}

\author{
  Chonghuai Ma \\
  Photonics Research Group\\
  Ghent University - imec\\
  Technologiepark 126, 9052 Gent, Belgium\\
  \texttt{chonghuai.ma@ugent.be} \\
  \And
  Floris Laporte \\
  Photonics Research Group \\
  Ghent University - imec \\
  Technologiepark 126, 9052 Gent, Belgium\\
  \texttt{floris.laporte@ugent.be} \\
  \And
  Joni Dambre\\
  IDLab\\
  Ghent University - imec \\
  Technologiepark 126, 9052 Gent, Belgium\\
  \texttt{joni.dambre@ugent.be} \\
  \And
  Peter Bienstman \\
  Photonics Research Group \\
  Ghent University - imec \\
  Technologiepark 126, 9052 Gent, Belgium\\
  \texttt{peter.bienstman@ugent.be} \\
}

\begin{document}
\maketitle

\begin{abstract}
Using optical hardware for neuromorphic
computing has become more and more popular recently due to its efficient high-speed data processing capabilities and low power consumption. However, there are still some remaining obstacles to realizing the vision of a completely optical neuromorphic computer. One of them is that, depending on the technology used, optical weighting elements may not share the same resolution as in the electrical domain. Moreover, noise and drift are important considerations as well. In this article, we investigate a new method for improving the performance of optical weighting, even in the presence of noise and in the case of very low resolution. Even with only 8 to 32 levels of resolution, the method can outperform the naive traditional low-resolution weighting by several orders of magnitude in terms of bit error rate and can deliver performance very close to full-resolution weighting elements, also in noisy environments. 
\end{abstract}

\keywords{
Neuromorphic computing \and Silicon photonics \and Weight quantization}

\section{Introduction}
Machine learning is becoming ubiquitous in people's daily lives. It can achieve outstanding performance on a variety of tasks\cite{LeCun2015,Mnih2015,Krizhevsky2012}. However, the surge of the vast volumes of generated data is approaching the limits of the conventional Von Neumann architectures in electrical hardware, as Moore's law appears to be coming to an end. Optical neuromorphic computing\cite{Kristof14,Laporte18,Duport2012,Shen17,Lin2018}, with its high-efficiency high-speed data processing capabilities, is becoming more and more a viable choice as a hardware implementation for neural networks, especially for problems with high data volumes where the input is already in the optical domain. Integrated silicon photonics neuromorphic systems have proven themselves as potential options, especially since they can exploit CMOS fabrication technology and are therefore suited for mass production\cite{Rahim2018}.

\subsection{Photonic Reservoir computing}
\begin{figure}
  \centering
  \includegraphics[width=0.45\textwidth]{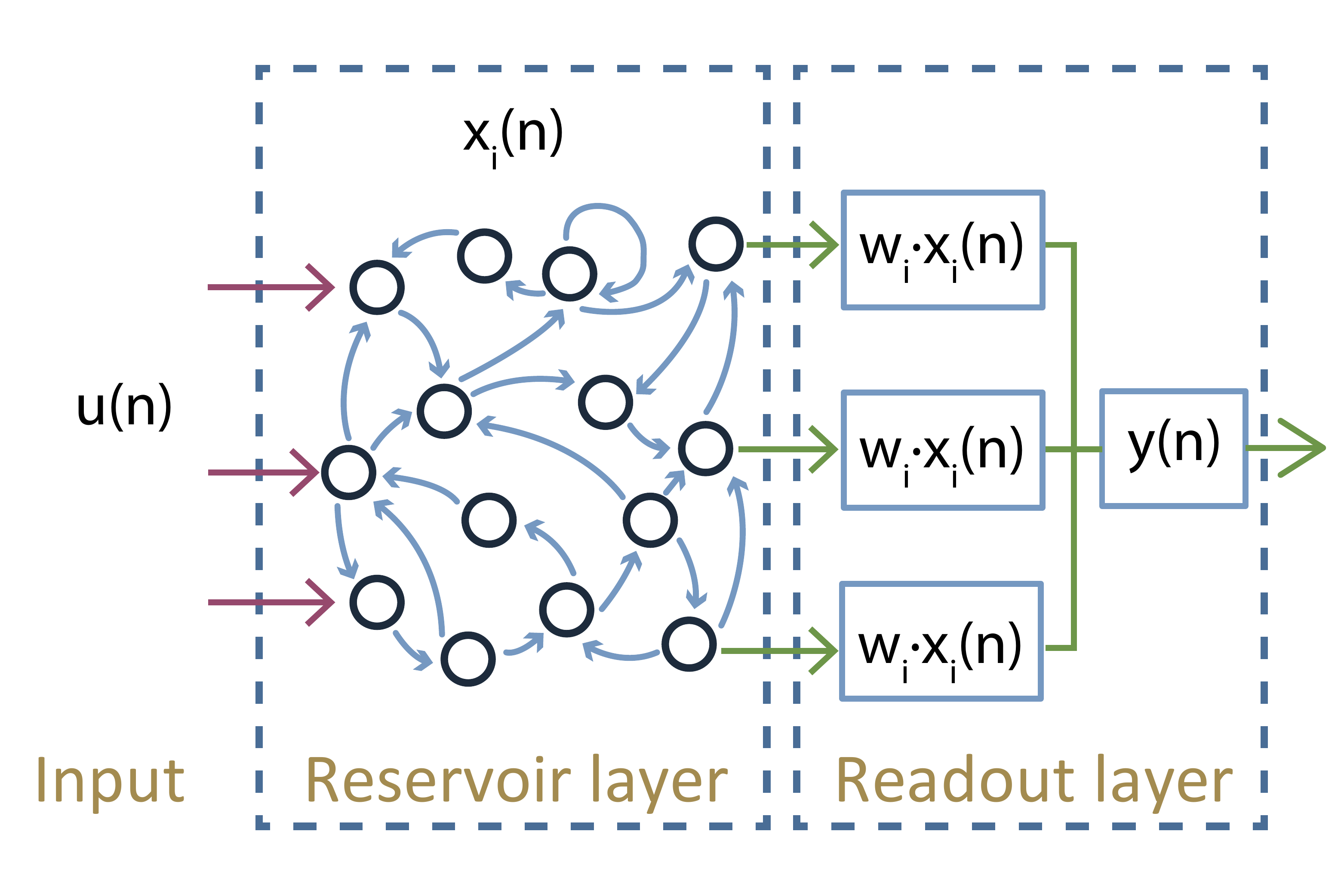}
  \caption{Reservoir computing consists of a reservoir layer, which is inherently an untrained RNN, and a readout layer that weights and linearly combines of output channels of the reservoir.}
  \label{fig:Reservoir}
\end{figure}

Reservoir computing (RC) is one of the machine learning paradigms \cite{Jaeger2004,Maass2002} whose relaxed requirements make it well suited for a hardware implementation. A reservoir is inherently a randomly initialized, untrained recurrent neural network (RNN), which acts as a temporal prefilter to transform a time dependent input into a higher-dimensional space, where it can be more easily classified by a linear classifier. Therefore, unlike RNNs, reservoir computing does not rely on optimizing the internal interconnection parameters of the network. Instead, it only optimizes the weights of the linear combination in the readout layer as shown in Figure~\ref{fig:Reservoir}. Because the performance of the reservoir is, within certain bounds, relatively insensitive to the exact internal weights of the reservoir, this technology is very interesting for a (photonic) hardware implementation, where fabrication tolerances are inevitable. Specific to photonic implementations are the so-called passive reservoirs, where the reservoir itself does not contain any nonlinearity, but where the photodetector in the readout (which converts a complex-valued light amplitude to a real-valued power intensity) provides the required non-linearity. Doing away with the need for internal nonlinearities reduces the power consumption. Photonic integrated circuits like these have been reported in \cite{Kristof14,Katumba15,Katumba17,Katumba18}, solving parity bit tasks, header recognition and telecom signal regeneration tasks at a data rates around 10 Gb/s. Laporte \textit{et al.} \cite{Laporte17,Laporte18} use photonic crystal mixing cavities as the reservoir and also could solve the XOR task on neighboring bits, with the potential of achieving much higher speeds, theoretically up to several hundreds of Gb/s.

In the meantime, other neuromorphic computing approaches are gaining prominence. E.g. optical feed-forward neural networks for deep learning are proposed by Shen \textit{et al.} \cite{Shen17}, 
which cascade integrated coherent optical matrix multiplication units embedded in programmable nanophotonic processors. The results show its utility for vowel recognition. As another approach, diffractive optical machine learning \cite{Lin2018} is achieved using diffractive elements as interconnection layers, which can solve various computational functions. A multiple-wavelength-based system is presented in \cite{DeLima2019}.

\subsection{Readout system}
\begin{figure}
  \centering
  \includegraphics[width=0.45\textwidth]{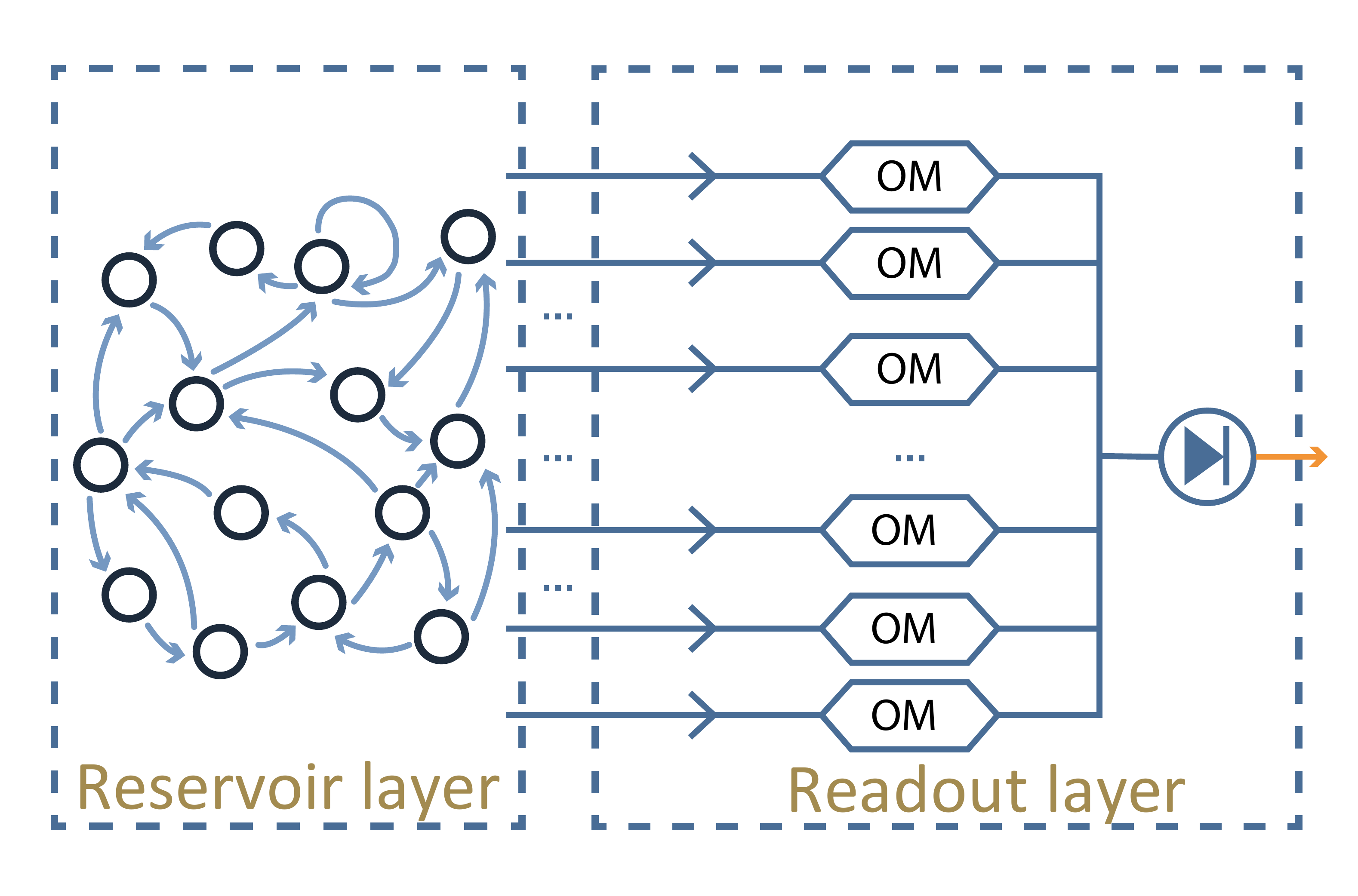}
  \caption{An integrated optical readout takes advantage of standard optical modulators (OM) to weigh the optical signals from the reservoir in both amplitude and phase with no latency and very low energy cost. The linear combination of the optical signals can be achieved using optical combiners.}
  \label{fig:Opt_rd}
\end{figure}

\begin{figure}
  \centering
  \includegraphics[width=0.45\textwidth]{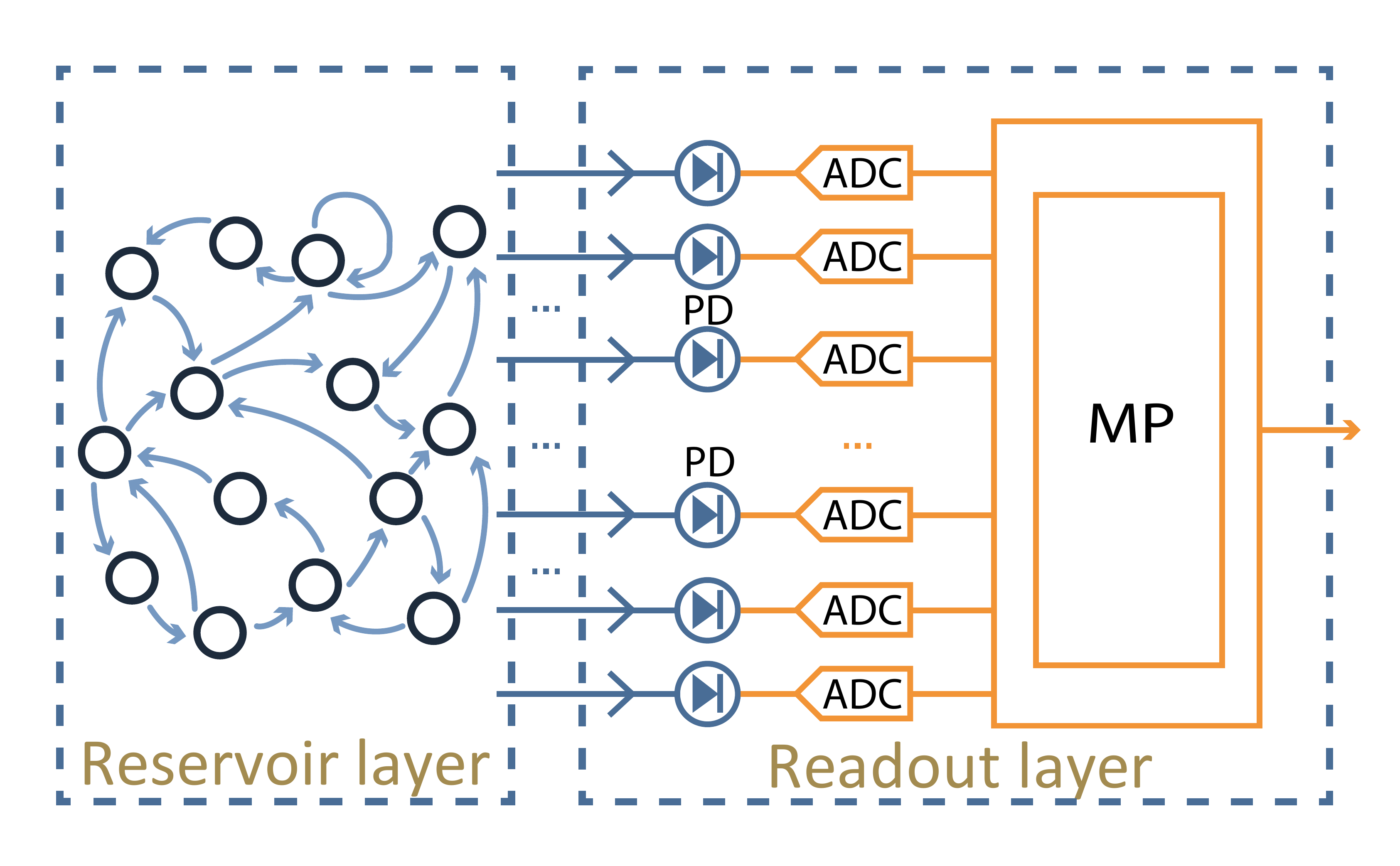}
  \caption{A Conventional integrated electrical readout requires an individual photodetector (PD) and Analog-To-Digital converter (ADC) for each output channel, as well as a microprocessor (MP) to perform the linear combination of the signals, which introduces unwanted latency and power consumption.}
  \label{fig:Elect_rd}
\end{figure}

To fully exploit a coherent optical neuromorphic computing circuit, thus benefiting from its high data rates and low energy consumption, an integrated optical readout to calculate the weighted sum (Figure~\ref{fig:Opt_rd}) is preferable. The
critical aspect for such an optical readout is that the weights are applied in the optical domain, instead of converting optical signals to electrical signals, and then applying the weights on the electrical signals using a microcomputer (referred as electrical readout schemes as shown in Figure~\ref{fig:Elect_rd}). Apart from speed, latency, and power efficiency advantages, such a fully optical readout system can also perform weighting on phase and amplitude separately, which results in more degrees of freedom, and thus more computation capability. 

One issue of an optical readout is that there is no longer an explicit electronic observability for each of the states, since in such systems it is desirable to only have one single photodiode and high-speed analog-to-digital converter after the linear combination. This observability problem was dealt with by Freiberger \textit{et al.} \cite{Freiberger18}. Another problem is that, depending on the technology, the resolution of the optical weights could be much lower compared to applying weights in the electrical domain, which can easily reach 16 bit. In this paper, we will consider as an example an implementation of optical weights in Barium Titanate (BTO) \cite{Abel16}, which has the critical advantage of being non-volatile, consuming nearly zero power while weighing the signal. The big drawback is that the resolution of the refraction index change limits the resolution to around 10 to 30 levels, coupled with some inevitable noise due to drifting of the elements. 

\subsection{Weight quantization}
Network quantization is a trending topic for deep learning researchers and developers, since the demand for applying machine learning tasks on low power consumption systems, e.g. mobile devices, has been increasing drastically\cite{Jacob2018}. From a certain point of view, a part of the problem that this paper is trying to solve is weight quantization. However, the network quantization concept in the context of deep learning and the weight quantization concept in coherent optical neuromorphic hardware systems are not the same. The differences are reflected in the following four aspects. 

First, for deep learning network quantization in the electrical domain, the resolution drop is usually from 32-bit floating-point to around 8-bit floating-point or 16-bit fixed-point with the aim of reducing memory and power consumption. In contrast, weight quantization in optical hardware could be way more severe, with resolutions going down to around 10 to 30 levels.

Secondly, the deep neural networks have a substantial number of trainable parameters; some of them may reach tens of millions, leading to a significant redundancy in deep learning models. For optical systems like passive reservoir computing chips, the number of trainable parameters in the readout system is typically in the hundreds or fewer, leading to a reduced parameter redundancy and therefore a potentially much more severe impact on the performance by the weight quantization. 

The third difference is that for quantization in a CPU- or GPU-based deep network, there is no noise on the quantization levels, whereas, in optical weighting elements, noise and drift are inevitable, and can sometimes be significant. 

Lastly, integrated optical elements are often not able to deliver a weight of zero (i.e. fully blocking the light signal). This is quantified in the extinction ratio of the weight, and impacts the performance of the system as well.

In such a context, training a readout system for coherent optical neuromorphic computing systems so that the performance is less sensitive to the quantization and the noise of the weights is of great importance. It is precisely this problem we want to tackle in this paper.

The rest of this paper is structured as follows. In Section 2, we introduce our modeling method of the optical weighting hardware, in amplitude and phase, by taking into consideration the number of quantization levels, the extinction ratio of the elements, and the noise after quantization. The explorative quantization retraining technique used in this work will also be highlighted. In Section 3, we show our simulation results based on our quantization method with respect to the different hardware parameters mentioned in Section 2. We also discuss the quantization result on various tasks, from easy ones to hard ones. 

\section{Methods}

In this part, we introduce the concept of our quantization method and present the implementation details like the discretization model, the type of regularization and the training procedure. 

\subsection{Explorative quantization weight selection}

The explorative quantization weight selection in this work is inspired by methods that have been used in deep learning quantization. Typically, after a full-precision model has been trained, a subset of weights is identified to be either pruned \cite{Han2015} or kept fixed \cite{Zhou2017}. The other weights are then retrained in full precision and requantized. If necessary, this step can be repeated in an iterative fashion, retraining progressively smaller subsets of the weights in order to find the most optimal and stable solution.

A crucial part of these methods is selecting a subset of weights to be left fixed or to be pruned. Random selection of weights is not a good idea, because there is a high probability of eliminating 'good' weights that convey important information. \cite{Han2015}\cite{Zhou2017} tackle this problem by choosing the weights with the smallest absolute value.

This is reasonable in deep learning models, since the millions of weights can provide enough tolerance when it comes to accidentally selecting the 'wrong' weights. However, in the readout systems we are investigating here, we have much fewer weights and a much more limited resolution with severe noise. In this case, the absolute value will not provide enough information, as a combination of many small weights could be important in fine-tuning the performance of the network. This will lead to a risk of a huge accuracy loss when specific 'wrong' connections (that are more sensitive to perturbations) are chosen to be retrained.

Instead, we adapt a different (albeit more time-consuming) approach, where after quantization, we compare several different random partitions between weights that will be kept fixed and weights that will be retrained (in full precision) and requantized. By comparing the task performance for these different partitions, we are able to pick the best one.

To get the best results, we conduct the procedure above in an iterative way. With each iteration, the ratio of fixed weights increases, starting from an initial value of $0.5$. In each iteration step, we evaluate 20 different random weight partitions. We typically perform 4 iterations, each time increasing the ratio of fixed weights by a factor of two.

\subsection{Quantization of the optical weighting elements}

Here, we describe in more detail what model has been used for the quantization procedure itself.
We aim to provide a high-level model for the weighting elements (both amplitude and phase)
that is as generic as possible, without being tied to any specific hardware
implementation details. Rather, in the quantization process we take three 
major aspects into account: extinction ratio, resolution and noise.

\paragraph{Amplitude quantization.}

The readout systems that we model here are supposed to be for an optical system
without amplification. In this case, the largest weight of the amplitude will be 1 (light passing through 
without being affected) and the lowest weight will be 0 (light is fully blocked 
by the weighting elements). However, in a realistic optical weighting element, 
the fully blocked weight '0' is often not achievable. Most elements 
have an extinction ratio describing how much optical power it can block compared 
to the maximum power that can pass through. For example, given 
an an extinction ration of 10 and a maximum weight of 1, the minimum weight will be 0.1.

\begin{figure}
  \centering
  \includegraphics[width=0.50\textwidth]{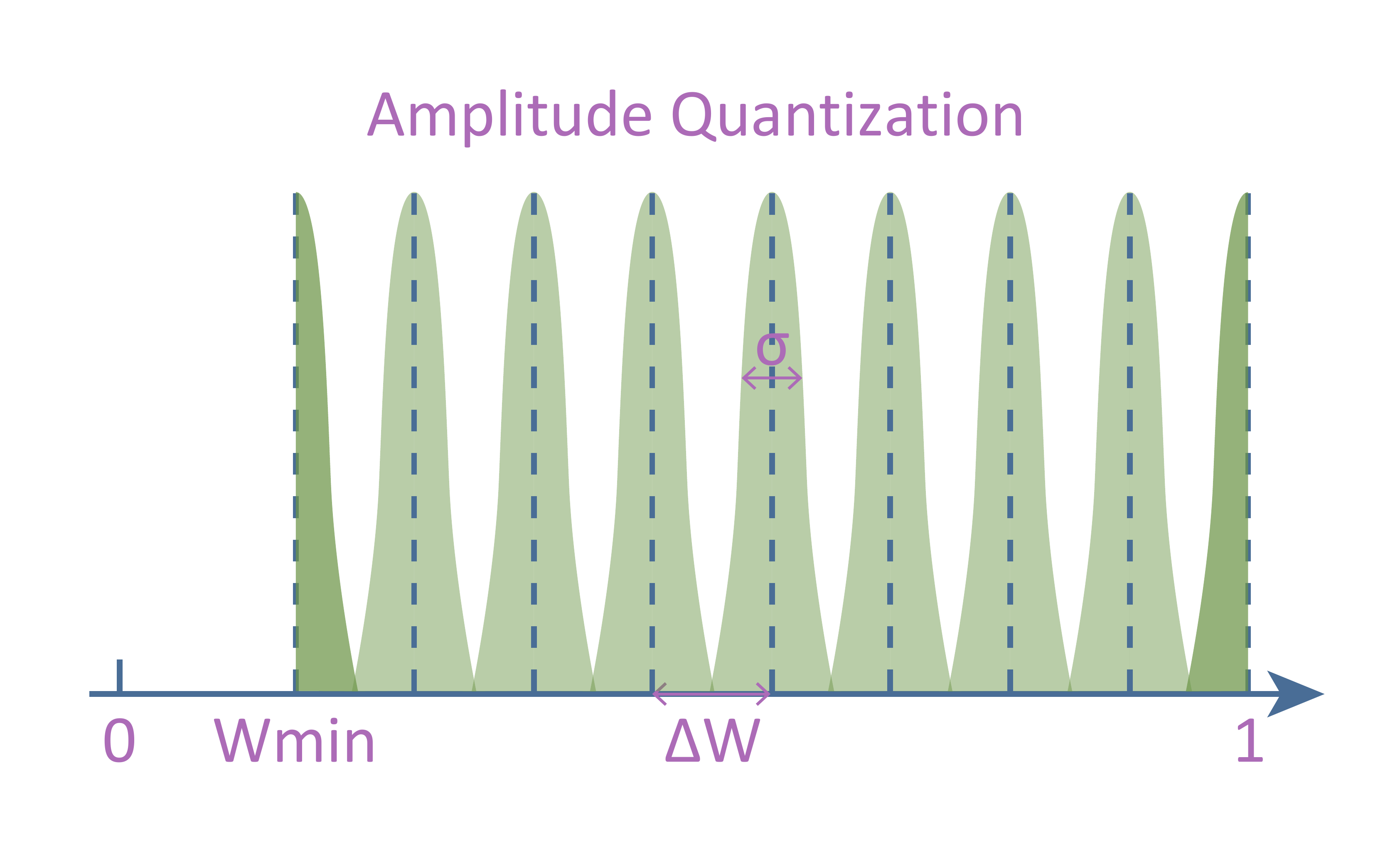}
  \caption{An illustration of amplitude quantization model for a general optical weighting element. The green shadowed areas represent drifting noise probability distribution.}
  \label{fig:Am_quiantize}
\end{figure}

The resolution of the weights is characterized as a number of bits or number of levels. All the possible weighting levels are uniformly distributed within the same interval, taking into account the minimum weight because of the extinction ratio. In more detail (Figure~\ref{fig:Am_quiantize}, the minimum weight the readout system can reach is given by:
$$ w_{min} = 1/\text{Extinction ratio}$$

After quantization, weights that are below this level will be rounded up to this value.

The distance between two adjacent weighting values is given by:
$$ \Delta w = (1 - \text{Extinction ratio}) / (N - 1) $$

where N is the number of available weighting levels, determined by the resolution of the optical element.

Noise and drift are of course very dependent on the details of the hardware used, but
for our purposes, we will abstract the noise as a Gaussian distribution. The noise level numbers we give in the following section are based on the standard deviation $\sigma$ of the Gaussian distribution as follows:
$$ \text{Noise level} = \sigma / \Delta w $$

\paragraph{Phase quantization.}
In the phase quantization process, the extinction ratio issue will not play a role, and there will
only be resolution and noise. The quantization will be implemented between 
$[0, 2\pi)$. All the levels are also evenly distributed. Noise is also modeled as Gaussian distribution.

\subsection{Regularization}
The regularization method used in this work is L2 regularization. In the photonic hardware readout system, there is no significant redundancy of connections or weights, and each weight is realized by a real waveguide in which the optical signals are passing through. It is not practical to 'cut' or 'prune' any hardware optical connections. Therefore we choose L2 over L1 regularization since we do not want to waste any readout waveguide L1 tends to result in zero weights. Choosing the best regularization parameter is crucial to prevent overfitting and to help the system to be robust against noise and quantization \cite{Hansen1987}. 

\subsection{Training procedure}
The original idea of reservoir computing is to use linear (ridge) regression to train the weights of a readout system. This saves a lot of training time compared to using gradient descent, as weighing the reservoir output signals is just a linear regression and can be calculated explicitly. In a coherent photonics reservoir however, the outputs of the reservoir are coherent optical signals containing amplitude and phase information, i.e. complex numbers. In the readout system, these optical signals are mixed together and interfere with each other to produce a final optical signal injected into a photodetector. Therefore the output of such a readout system is a single electrical signal containing only intensity information. This is a nonlinear conversion from a complex-valued optical signal with an amplitude and phase to a purely real intensity . To incorporate such a nonlinear conversion, we can no longer use the one-step solution based on the Moore-Penrose inverse of linear ridge regression, since there is no unique way to choose a phase of the signal before the photodiode that gives rise to the desired amplitude after the photodiode. Therefore, in our learning phase, we optimize the weights with gradient descent, explicitly taking the nonlinear photodiode into account. Using this method gives us more robust results and increases the computation capability of our reservoir system.

Another decision we made is to use the Mean Squared Error (MSE) as a loss function to optimize the optical weights. One could argue to use logistic regression instead, because the tasks we're trying to solve are all classification tasks. However, to build an optical neuromorphic computing system, we have to consider the hardware implementation of each signal processing step. In our hardware system, the output electrical signal from the final photodetector will be used as the final prediction signal without further signal processing. Using logistic regression would assume another layer of nonlinear calculation, the sigmoid function. This process would result in a more complex integration technology and induce more latency and energy cost.

\section{Experimental results}

In this section, we present simulation results for the performance of the proposed method on two tasks: header recognition and a boolean XOR operation. The bit sequences will be encoded to optical signals before the simulation. The architecture we choose for thses tasks is the 4-port swirl reservoir network \cite{sackesyn2018enhanced} with 16 nodes. We train all the tasks on fully integrated passive photonic reservoirs with optical integrated readout systems. We treat the full resolution weighting performance with weighting noise as a baseline, and also compare our method with the performance of direct quantization. 

The simulation consists of two parts. The first one is to generate the reservoir signal. We use Caphe\cite{Fiers2012} to simulate the reservoir circuit with subsampling of 20 points per bit on the intensity-modulated input signal. The delays in the photonic reservoir are optimized for an input signal speed of 32 Gbps. The response of the photonic reservoir will be the complex output signals from each of the nodes, consisting of amplitude and phase information. The second part is the  training and quantization of the weights using our proposed method. We use PyTorch\cite{paszke2017automatic} to implement this. 

\subsection{Header recognition}
The header we use in the header recognition task is a 4-bit header '1101'. We want the final readout signal to be '1' whenever the chosen header appears in the input signal, and '0' otherwise. The reservoir architecture is a $4\times4$, 16 node reservoir. Although header recognition can be solved in a multi-class classifier, the optical readout hardware we use here is not suited for multi-channel optical output. As mentioned, we use an MSE loss function with L2 regularization and train our models with gradient descent useing Adam \cite{Kingma2014}.

We now discuss the influence of resolution, noise and extinction ratio on the performance.

\begin{figure}
  \centering
  \includegraphics[width=0.95\textwidth]{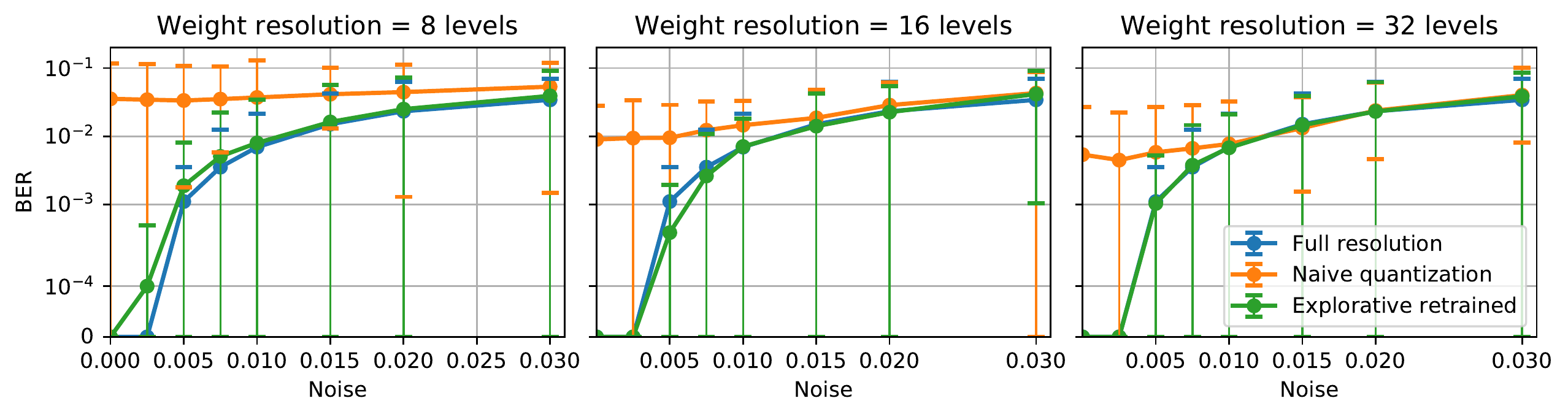}
  \caption{Performance comparison of three different weighting resolutions, 8 levels (top), 16 levels (middle), 32 levels (bottom) for the 4-bit header recognition task over different noise levels. The blue curve represents the performance of full-resolution weights; the orange curve represents naive quantization weights; the green curve for explorative retrained quantization weights.}
  \label{fig:HR_noises}
\end{figure}

\subsubsection{Noise}
Noise has a very significant influence on performance. Figure \ref{fig:HR_noises} shows the performance evaluation of the Bit Error Rate (BER) at different noise levels at 8 levels, 16 levels and 32 levels resolution on the 4-bit header recognition task. The top figure with 8 levels resolution shows that overall, the quantized weights obtained by our explorative retrain method consistently provide performance levels very close to the full resolution weights. For more substantial noise levels of the optical weighting elements as shown in the far right part of the figure, the BER increases significantly, which reflects the severe impact that the drifting noise of the optical weighting elements brings to system performance, even for a full resolution system. Apart from those situations of large noise levels, the retrain method is performing well throughout and gives several orders of magnitude better BER than the naive quantization of the weights where the readout is trained and quantized only once. When there is no noise on the weights, the retrain method gives almost the same performance on the limited resolution readout system compared to a full resolution readout system. This result is rather surprising, given that the resolution here is only 8 levels.
 
For higher resolutions, namely 16 levels and 32 levels, the explorative retraining method provides even closer performance to the full resolution readout systems. As the resolution increases, the direct quantization method also achieves lower bit error rates, however still not comparable to the full resolution and explorative retrained weights. With high noise involved on the other hand, the three sets of weights are giving a similar performance (especially at 32 levels resolution all three results are entirely indistinguishable), The reason is that, as the resolution increases, the interval between two adjacent weighting level is in the same range of the variance of the noise. These three weighting methods will provide similar weighting strategies statistically.

Finally, we want to note that, in the middle figure where the readout resolution is 16 level, the retraining method sometimes gives a small performance improvement compared to the full resolution weights, which is very interesting. This is mainly because, during the retraining, the network is continuously choosing weights to optimize so that the system becomes more tolerant to quantization. This procedure sometimes results in finding a better local minimum in the cost landscape. However, this advantage is marginal and only happens occasionally.

\begin{figure}
  \centering
  \includegraphics[width=0.95\textwidth]{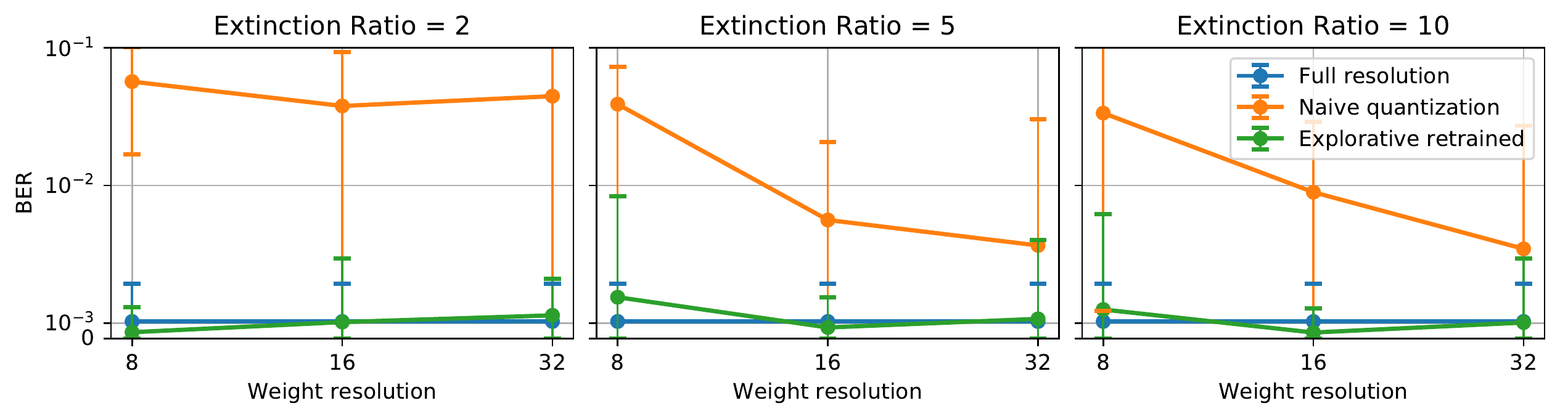}
  \caption{Performance comparison of three different extinction ratio, 2 (top), 5 (middle), 10 (bottom) for the 4-bit header recognition task over different resolutions. The blue curve represents the performance of full-resolution weights; the orange curve represents naive quantization weights; the green curve for explorative retrained quantization weights.}
  \label{fig:HR_Reso}
\end{figure}

\subsubsection{Extinction ratio}
Figure \ref{fig:HR_Reso} shows that for directly quantized weights, a higher resolution always gives better BER regardless of the extinction ratio of the weighting components. However, when the extinction ratio is higher, the performance from the 32 levels naive quantized weights is closer to the full resolution weights. This result is intuitive since higher extinction ratio means the weight range is more preserved after quantization, and when the extinction ratio is low, more initially lower weights will be rounded up to the lowest weighting value that the extinction ratio allows. 

In contrast, the retraining method gives constantly better performance because the weights are continually evolving to adapt to the weighting range that the extinction ration defines. The performance levels are always very close to those of the full resolution, with a limited impact of extinction ratio. 

\subsection{4-bit delayed XOR}
We chose 4-bit delayed XOR (i.e. calculating the XOR of the current bit and 4 bits ago) as our second task. The reason is that XOR is a more nonlinear task compared to the header recognition task, so it is interesting to see how it behaves under weight quantization situations. 

\begin{figure}
  \centering
  \includegraphics[width=0.95\textwidth]{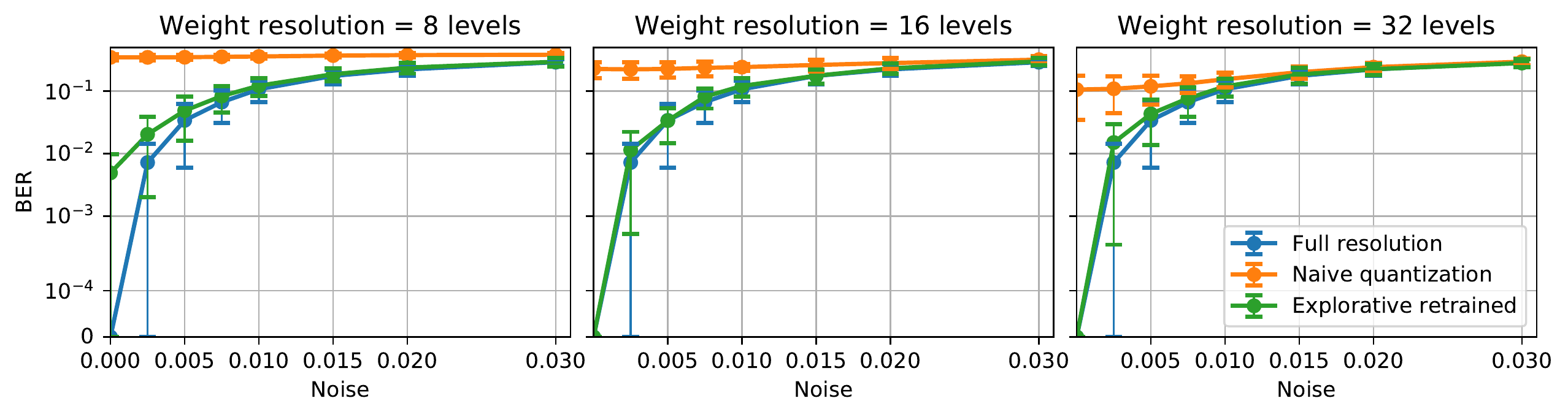}
  \caption{Performance comparison of three different weighting resolutions, 8 levels (top), 16 levels (middle), 32 levels (bottom) for the 4-bit delayed XOR task vs different noise levels. The blue curve represents the performance of full-resolution weights; the orange curve represents naive quantization weights; the green curve for explorative retrained quantization weights.}
  \label{fig:XOR_noises}
\end{figure}
\begin{figure}
  \centering
  \includegraphics[width=0.95\textwidth]{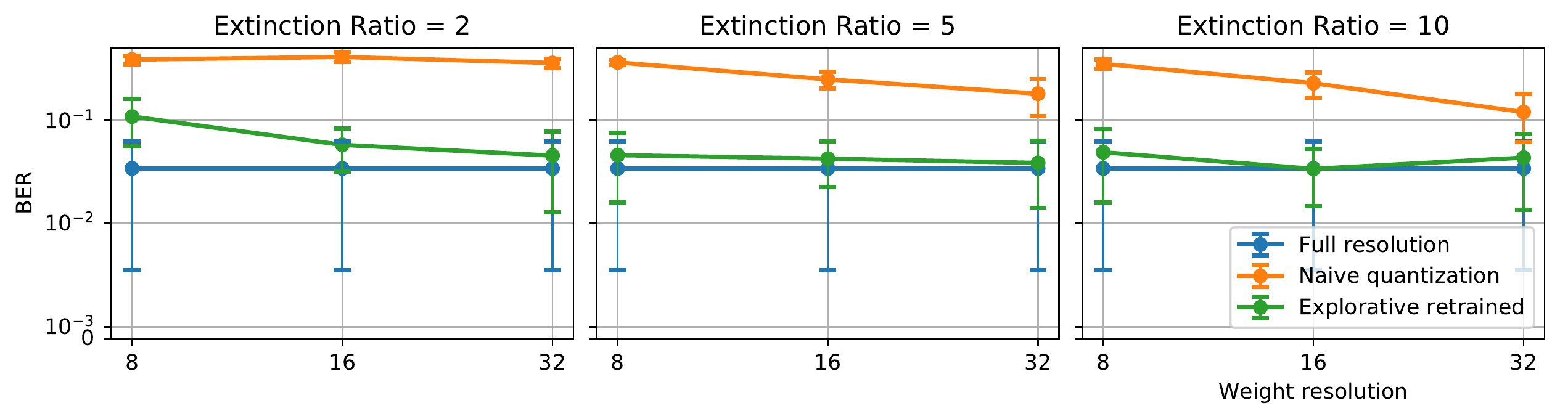}
  \caption{Performance comparison of three different extinction ratios, 2 (top), 5 (middle), 10 (bottom) for the 4-bit delayed XOR task over different resolutions. The blue curve represents the performance of full-resolution weights; the orange curve represents naive quantization weightsp; the green curve for explorative retrained quantization weights.}
  \label{fig:XOR_Reso}
\end{figure}

Figure \ref{fig:XOR_noises} presents the performance of our quantization retraining method on low-precision weights (8, 16, 32 levels) at different noise levels. From the blue curve, it can be seen that 4-bit delayed XOR is indeed a harder task to tackle since even with full precision, the BER increases significantly with a small amount of noise. Moreover, the task is also very sensitive to low precision weights, as shown from the naive quantization weights (orange curves), at the 8 levels resolution, the tasks is unsolvable even without any noise involved. In the meantime, we also observe that our explorative retrained weights are capable of providing very close performance to full resolution weights, except for 8 levels resolution, due to the extra nonlinearity requirement of the task, retrained weights find themselves hard to follow. But still, naive direct quantized weights are severely outperformed by our explorative retraining method at the overall spectrum of the different noise levels.

The performance dependence on resolution and extinction ratio is shown in Figure \ref{fig:XOR_Reso}. The noise level here is $0.005$. Similar to the header recognition task, a higher extinction ratio provides better direct quantization performance. Weighting elements with an extinction ratio of 2 are not able to provide enough weighting range for direct quantized weight to deliver workable performance regardless of the weighting resolution. The explorative retrained method is also affected at the 8 levels resolution. When the extinction ratio is 5 and 10, the quantization retraining method delivers performance close to the full resolution, from which we can draw a statement that an extinction ratio of 5 is sufficient for this task.

\section{Conclusion}
In this paper, we addressed the limited weight resolution in an integrated all-optical readout system for photonic  reservoir computing systems. A high-level general modeling for realistic integrated optical weighting elements enables us to characterize the influence of the weighting resolution, drifting noise, and extinction ratio on the system performance. Our proposed explorative retraining method focused on identifying the best weights to be retrained. It is shown that in situations where both the number of output channels and the weighing resolution is extremely limited, our proposed method still delivers performance very close to that of full resolution weights over a large span of weighting noises.

\section*{Acknowledgment}

This research was funded by Research Foundation Flanders (FWO) under Grant 1S32818N, the EU Horizon 2020 PHRESCO Grant (Grant No. 688579) and the EU Horizon 2020 Fun-COMP Grant (Grant No. 780848).

\bibliographystyle{unsrt}  
\bibliography{Chonghuai_Ma}






\end{document}